# Effect of high temperature service on the complex through-wall microstructure of centrifugally cast HP40 reformer tube

Thibaut Dessolier[1,*], Thomas McAuliffe[1], Wouter J. Hamer[2], Chrétien G.M. Hermse[3] & T. Ben Britton[1]

[1] Department of Materials, Imperial College, Exhibition Rd, London, SW7 2AZ, UK

[2] Shell Global Solutions International B.V., Grasweg 31, 1031 HW, Amsterdam, The Netherlands

[3] Shell Nederland Chemie B.V., Chemieweg 25, 4782 SJ Moerdijk, The Netherlands

* corresponding author: t.dessolier@imperial.ac.uk

## Abstract

Centrifugally cast reformer tubes are used in petrochemical plants for hydrogen production. Due to the conditions of hydrogen production, reformer tubes are exposed to high temperature which causes creep damage inside the microstructure. In this study, two different ex-service HP40 alloy reformer tubes which come from the same steam reformer unit have been compared by microstructural characterisation performed at a range of length scales from mm to µm. Analyses performed by EBSD (Electron Backscatter Diffraction), EDS (Energy Dispersive X-ray Spectroscopy) and PCA (Principal Component Analysis) show that both tubes have similar microstructural constituents, with the presence of an austenitic matrix and $M_{23}C_6$, G phase and $M_6C$ carbides at the grain boundaries. Even if both tubes have a similar microstructure, one tube due to it localisation inside the steam reformer unit presents a region with more micro cracks which may indicate that this tube have accumulated more creep damage than the other one.

## 1. Introduction

The performance of engineering alloys underpins their economic utility in modern chemical engineering plants. Understanding microstructural evolution, and creep, of components can maximise their use, by providing insight to create models that predict component failure and thereby reduce unscheduled plant downtime and outage.

Centrifugally cast alloys [1] are generally used in petrochemical plants as reformer tubes for hydrogen production by steam reforming. Reformer tubes are located inside a furnace, where hydrocarbons and steam flow inside and react to produce hydrogen. This typically is performed within the temperature range of 700-1000°C and at a pressure of 2 to 4 MPa [2–4] These temperatures can cause a change in the microstructure, and when under mechanical load creep damage may accumulate. The creep is driven by a combination of temperature, time and hoop stress and can continue until tube failure [4,5].

In these components, proprietary iron-nickel based alloys similar to the generic HP type alloys (≈ Fe - 35Ni - 25Cr – 1.5Nb - 0.4C wt% plus minor other elements) are used as they have a good resistance against high temperature creep and carburization. Long tubes are formed using centrifugal casting and this results in a complex through-wall microstructure, with equiaxed grains at the tube inside diameter (ID) and columnar grains towards the outside diameter (OD) together with elemental segregation and changes in local microstructure based upon the time and temperature history. Based on prior work [7–11], in the as-cast condition standard HP alloys are composed of an austenitic matrix with some niobium rich carbides (identified as NbC) and chromium rich carbides (identified as $M_7C_3$ and $M_{23}C_6$) at the grain boundaries. De Almedia Sores et al. [8] have demonstrated that $M_{23}C_6$ carbides are preferred for higher Nb content (Nb > ≈ 2% wt), or based on the work of Yan et al. [12] due to addition of tungsten. These carbides are important in these components as they underpin mitigation against creep accumulation by inhibition of grain boundary sliding and they provide direct strengthening by obstructing dislocation motion [3,7,13,14].

Examination of material taken from service indicates that the carbides may transform within the microstructure [4,11 and 15–17]. Some studies [8,9,15,18,19] indicated that $M_7C_3$ carbide transformed into $M_{23}C_6$ carbide due to carbon diffusion from the austenite matrix at the grain boundary to the existing chromium rich carbide phase [16,18,20], and concurrently there were $M_{23}C_6$ carbide present inside the grains. It is likely that this is favourable, as this precipitation leads to a growth of the primary chromium network and consequently increases the creep resistance as described by Tancret et al. [21].

Further to these transformations, high temperature aging results into the partial transformation of NbC carbide (due to the instability of NbC precipitates between 500 and 1100°C) to a nickel-niobium-silicon rich intermetallic commonly known as G phase [11,15,16,22,23]. The G phase is often located on the grain boundaries, which is thought to be the prior site of NbC precipitates. This G phase composition is mostly reported as $Ni_{16}Mn_6Si_7$ or $Ni_{16}Nb_6Si_7$ [10,18,19,22,24,25,26] with a face centred cubic (FCC) structure. Within the HP alloy, the addition of Nb and Si has been found to facilitate G phase formation [8,11,18,19,22] and the G-phase can be reduced when minor elements such as Ti and W are co-located in the NbC [12,17,18].

There is an alternative phase transformation of the NbC carbides reported by [27 and 28], where NbC decomposes into nickel-niobium-silicon-chromium rich phase also known as $M_6C$ carbide or also reported as η phase [29 and 30]. Buchanan et al. [10] have noticed that for a similar alloy the presence of either G phase or $M_6C$ carbide while [16 and 30] reported the presence of both phases in the same alloy. Similar as the G phase, the $M_6C$ carbide is generally located on the grain boundaries and Nb and Si can facilitate $M_6C$ carbide formation [30, 28, 31]. Moreover, $M_6C$ carbide has a diamond cubic structure with the composition $Cr_3Ni_2SiC$ [10, 30, 29]. Furthermore, the Cr/Ni ratio of the phase was found to be appreciably higher (~1) in comparison to those reported for the G phase (~0.1) [30]. Tancret et al. [18] noted that the transformation of NbC to G phase or/and $M_6C$ carbide is function to the temperature, time and silicon content.

Table 1 recapitulates all the phases that could be present inside HP series alloy after aging. Prior work indicates that the $M_6C$ carbide is less common than the and G phase. However, Tancret and Buchanan [18 and 30] noticed that structure between G phase and $M_6C$ carbide are very similar and that differentiation between both is difficult and should merit further investigation. Table 2 presents the chemical composition of each carbide based on TEM EDX performed after ageing on HP alloys in literature.

| Phase | Crystal system | Space group | Lattice parameters – a, b, c (nm) | | | Ref |
|---|---|---|---|---|---|---|
| $\gamma$ | Cubic | $Fm\bar{3}m$ | 0.36 | 0.36 | 0.36 | [10] |
| $M_{23}C_6$ | Cubic | $Fm\bar{3}m$ | 1.06 | 1.06 | 1.06 | [10] |
| NbC | Cubic | $Fm\bar{3}m$ | 0.45 | 0.45 | 0.45 | [10] |
| G phase $Ni_{16}Nb_6Si_7$ | Cubic | $Fm\bar{3}m$ | 1.12 | 1.12 | 1.12 | [10] |
| $M_6C$ carbide $Cr_3Ni_2SiC$ | Cubic | $Fd\bar{3}m$ | 1.06 | 1.06 | 1.06 | [30] |

**Table 1: Synthesis of the phases which can be present after aging or for ex-service HP series alloy.**

| % at | Ni | Cr | Fe | Nb | Si | V | Ref |
|---|---|---|---|---|---|---|---|
| $M_{23}C_6$ | 10 | 36 | 7 | 1 | 4 | / | [22] |
| NbC | / | 4 | 1 | 57 | 2 | / | [22] |
| G phase $Ni_{16}Nb_6Si_7$ | 49.6 | 6.6 | 5.9 | 16.8 | 21.1 | / | [30] |
| $M_6C$ carbide $Cr_3Ni_2SiC$ | 29.9 | 30.6 | 5.2 | 12.2 | 14.8 | 1.3 | [30] |

**Table 2: EDX composition (% at.) of each element based on literature.**

In this study, machine learning based correlated EBSD and EDX analysis using principle component analysis and pattern matching (based upon the method introduced in prior work of [32]) is used to more reliably classify each of the carbide types and link this to longer length scale characterisation based upon segmentation of backscatter electron micrographs.

The samples involved are from the centrifugally cast HP40 type. The samples from two ex-service reformer tubes which have been in service under differing conditions are compared, providing the ability to explore microstructure evolution in these alloys for different creep life. In this article, microstructure characterisation is performed at a range of length scales from mm to µm. The presence and distribution of different grain structures and carbide structures are explored and evaluated with respect to creep life and the presence of micro-cracks.

## 2. Materials and methods

Two ex-service HP40 alloy reformer tubes are characterised, and these were provided by Shell. Both tubes have been in service for almost 30 years (≈ 250 000 hrs of production) and were used in steam methane reforming (SMR) service to produce a hydrogen mixture ($CH_4 + H_2O \rightarrow CO + 3\,H_2$), where the hydrogen product was purified by pressure swing absorption and ultimately has a >99% $H_2$ purity (in mole %) at the end of the process.

During the hydrogen production, the operating temperature for the nickel catalyst is ≈ 850°C. As the process is endothermic, the temperature along OD can reach ≈ 920°C. While the tubes were taken from the same steam reformer unit at the same time, however, tube B was subjected to more stress and temperature (due to the geometry of the reformer) and therefore has crept more.

Centrifugal casting was used to produce tubes of better 4 or 5 meters long and these are joined together using electron beam welded to form the final tube with a length of 12 to 14 meters. In our case both tube samples were 1 meter long and have been sectioned at ≈ 1 m from the bottom of the furnace. Tubes A and B have an OD equal to 130 mm and an ID equal to 100 mm. To determine the tubes composition, EDS was performed at the mid wall section of each tube. This area (scan size of ~ 0.8 x 0.5 mm) can be considered as the most representative for the original material composition. Table 3 presents average composition of HP alloy in wt% from the literature while the composition of our tubes in wt and at% after EDS analyses.

Compare to literature information, both tubes here seem to have less nickel content (31.1 when the lower limit is 34 wt%) and more chromium content (28.8 when the upper limit is 27 wt%). The carbon content, though not measured, is expected to be similar, about 0.4 to 0.5 wt%. Apart from the elements mentioned below no significant amount of other elements were detected. Considering the EDS measurement error range, it is possible to advance than EDS result may correspond to the signature of HP40 alloy.

|  | C | Fe | Ni | Cr | Nb | Si | Mn |
|---|---|---|---|---|---|---|---|
| [1,3,4,8,22] | 0.3 - 0.5 | bal | 34 - 37 | 24 - 27 | 0.5 – 1.5 | < 2 | < 2 |
| Tube A (wt%) | n.m. | 36.6 | 31.1 | 28.8 | 0.7 | 0.9 | 1.7 |
| Tube B (wt%) | n.m. | 36.6 | 31.1 | 28.8 | 0.7 | 0.9 | 1.7 |
| Tube A (at%) | n.m. | 36.2 ± 1.5 | 29.3 ± 1.3 | 30.6 ± 1.2 | 0.4 ± 0.1 | 1.8 ± 0.1 | 1.8 ± 0.2 |
| Tube B (at%) | n.m. | 36.2 ± 1.5 | 29.3 ± 1.3 | 30.6 ± 1.2 | 0.4 ± 0.1 | 1.8 ± 0.1 | 1.7 ± 0.2 |

n.m. = not measured / bal = balance

Table 3: Average composition in (at.%) and (wt%) for both tubes after performing EDS with also the composition of HP40 based on the literature.

The tubes were sectioned using electrical discharge machining, then hot-mounted in bakelite (at ~190°C) and ground with SiC grinding papers, and ultimately polished with 0.04 µm colloidal silica to obtain a smooth surface. For Electron Backscatter Diffraction (EBSD) characterisation, each sample was polished by Broad Ion Beam Polishing for 1 hour, with a voltage of 1 kV and beam angle of 2°.

Imaging with second electrons (SE) and backscatter electrons (BSE) was carried out on a Zeiss Auriga SEM with a beam acceleration of 20 kV and a working distance (WD) of 10 mm. ImageJ was used for micrograph segmentation and analysis.

Analytical scanning electron microscopy was performed using a FEI Quanta 650 Scanning Electron Microscopy (SEM) equipped with a Bruker XFlash 6|60 detector for energy dispersive X-ray spectroscopy (EDS) analysis, and a Bruker eFlash[HD] EBSD camera was carried out for microstructural characterisation. Two types of EBSD measurement have been done. First for quick characterisation of the microstructure, in this case, EBSD measurements were performed using a beam acceleration of 20 kV and an aperture of 50 mm in high current mode, with eSprit 2.1. Each sample was analysed with an Electron Backscatter Pattern (EBSP) resolution of 200 x 150 pixels, a working distance (WD) of 15 mm, a step size of 2 µm and an EBSD camera exposure of 15 ms and a probe current of ~10 nA.

Higher spatial resolution phase analysis was also used to characterise in detail the carbide structure. In this case, EBSD measurements were performed using a beam acceleration of 30 kV and an aperture of 50 mm, with eSprit 2.1. Samples were analysed with an EBSP resolution of 800 x 600 pixels, a working distance (WD) of 15 mm, a step size of 100 nm and an EBSD camera exposure of 60 ms. For this analysis, EDS had was collected at 20 kV with an acquisition time of 180 s and with the sample tilted at 70°.

Correlative EDS and EBSD was performed using simultaneously acquired data (in a tilted configuration) and analysed used EBSD-EDS signal-weighted principal component analysis (PCA) approach following [32] method, and now available in AstroEBSD. In brief, the signals are weighted to balance the EDS and EBSD contributions prior to principal component analysis, which collects representative (combined) EDS and EBSD signals that strongly contributed to the maps and this increases the signal to noise within the representative signals. Using the method described in reference [32], a weight of 1 between EDS and EBSD signal was applied with a variance tolerance limit of 0.05 for the number of selected components prior to the rotation of the principal components and their subsequent analysis. The EDS data was captured with 2048 channels (i.e. a data resolution of 9.8 eV, while the detector has an energy resolution of ~126 eV). The EBSD data was background corrected and cropped to 300 x 300 pixels for the calculations.

The phases were distinguished based upon the representative EBSD patterns, using pattern matching with the refined template matching method developed by Foden *et al.* [33]. Classification was performed using EBSD pattern simulations of reference templates generated from representative phases within Bruker DynamicS.

## 3. Results and discussion

### 3.1. Tubes microstructure

#### 3.1.1. Global macrostructure

For each tube, the centrifugal casting process leads to an axially symmetric microstructure, with variations in structure along the radial axis. A section from both tubes is presented in Figure 1, with optical microscopy and EBSD orientation maps.

Tube A shows a typical microstructure, where the matrix phase (Ni-rich and FCC) includes very large columnar grains near the outer diameter, OD (region A-ii) that extends 15 mm from the OD to the inner diameter, ID. Tube A has a fraction of 70% of columnar and of 30% equiaxed grains along the tube section.

The grains are aligned with <001> along the radial, hoop and axial directions. In this section, the long axis of the dendrites extend along the radial direction, and these are ~2 mm wide. Within each large grain, there are low angle orientation misorientations and decorations of carbides between the dendrite arms. The boundaries between each dendrite (along the hoop direction) are wavy, indicating that the dendrites terminated on each other and that the nucleation of these grains started on the outer diameter of the tube.

Towards the inner diameter, there are smaller equiaxed grains with a grain size of ~25 µm (region A-i). There is a similar texture of the grains in this region, but it is significantly less sharp.

In Tube B, there is a similar columnar grain region (B-iv) near the outer diameter, followed by an equiaxed microstructure. However, in this tube there is a second columnar and equiaxed band towards the centre of the tube. The columnar grains in region B-ii are smaller than those found in region B-iv. This second band could be a result of a slight change in manufacturing.

As a note, manufacture of these tubes consists of boring of the centre of the tubes to remove slag and to obtain a controlled final inner diameter.

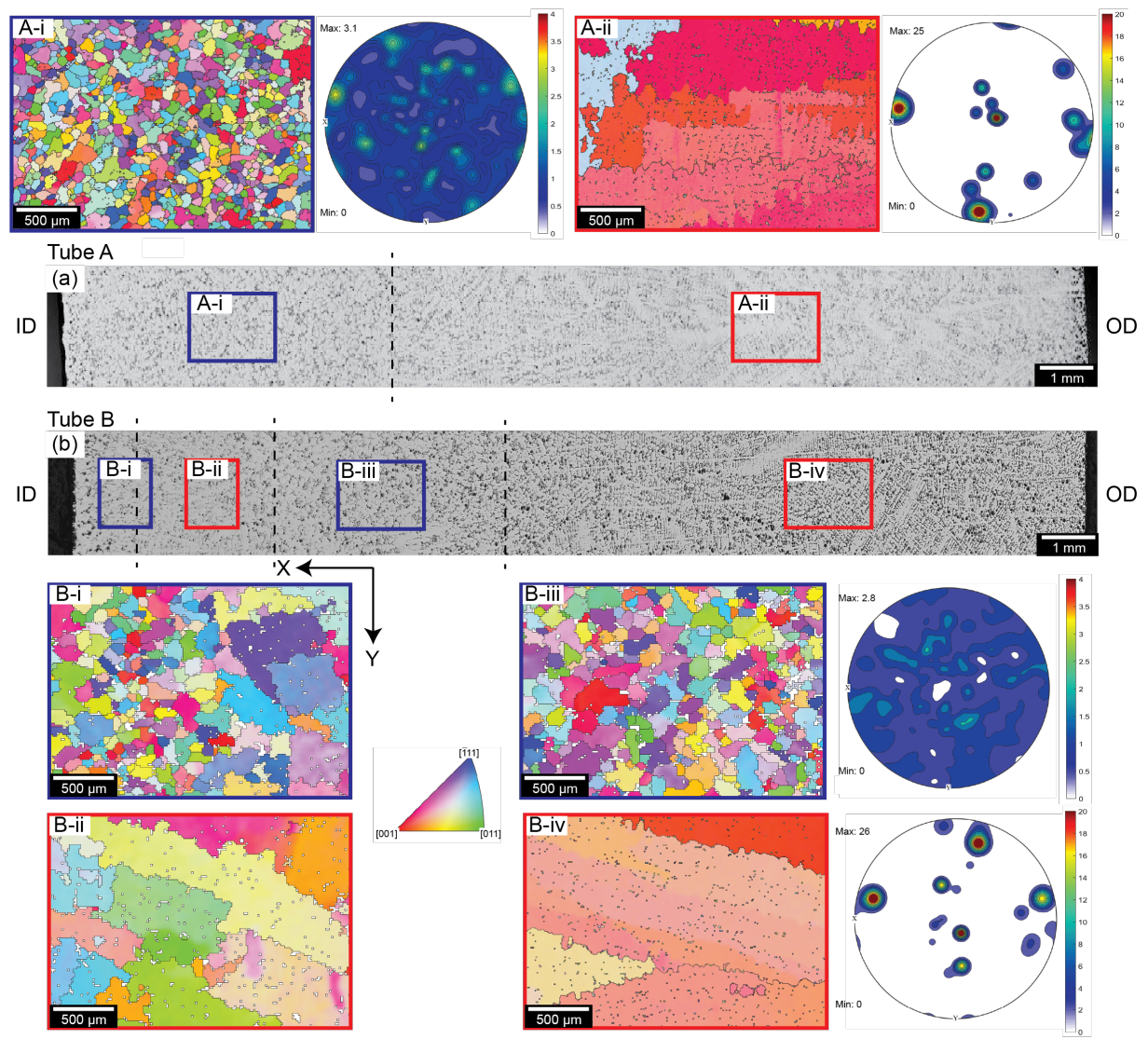

Figure 1: Comparison of the microstructure of both tubes with optical images of (a) tube A and (b) tube B. EBSD analyses were performed in several regions for both tube with Inverse Pole Figure map along the X direction. Regions A-i and A-ii correspond to the tube A while regions B-i, B-ii, B-iii and B-iv correspond to the tube B.

### 3.1.2. Carbides inside the microstructure

High magnification backscatter electron imaging was performed together with chemical analysis with EDS, and results are shown in Figure 2 and Figure 3. Carbides are found at grain boundaries and the intersection of the dendrite arms. The initial categorisation will be performed on images taken within the large dendrite/large-grain region.

The matrix is labelled 1 in Figure 2. Matrix has an intermediate grey level in backscatter imaging compare to carbide which have more contrast. EDS analysis in Figure 3 indicates the presence of Fe, Cr, Ni, and Si inside the matrix. This is likely an austenitic phase.

Three types of carbide (labelled 2-4 in Figure 2) were found in tubes A and B:

2 – carbides interlaced with other carbides (3 and 4) on the dendrite boundaries. They present as white in backscatter imaging, lighter than all other phases (i.e. the highest backscatter coefficient). EDS analysis indicates the presence of significant Nb and Si, less Cr and Fe than 1. There is high Ni content, but this may be a matrix contribution. The EDS spectrum does not correspond to a NbC spectrum as report by [30 and 34], and seems rather similar to the G phase.

3 – carbides interlaced with other carbides (2 and 3) on the dendrite boundaries, as large carbides extending from this interlaced carbide structure and also within the grain interior. Within the grain interior, these carbides have an irregular shape. They are darker than the matrix (i.e. a lower backscatter coefficient). EDS analysis indicates the presence of significant Cr, and limited Nb, Si, Ni and Fe. The EDS spectrum looks consistent with $M_{23}C_6$ spectra found by [29, 30, 34].

4 – Light grey carbides often found on the dendrite boundaries. They have an intensity very close to the intensity of the matrix (i.e. a similar backscatter coefficient). EDS analysis indicates the presence of Si, V, Nb, Cr and Ni. This is likely that the spectra correspond to the $M_6C$ carbide [30 and 35].

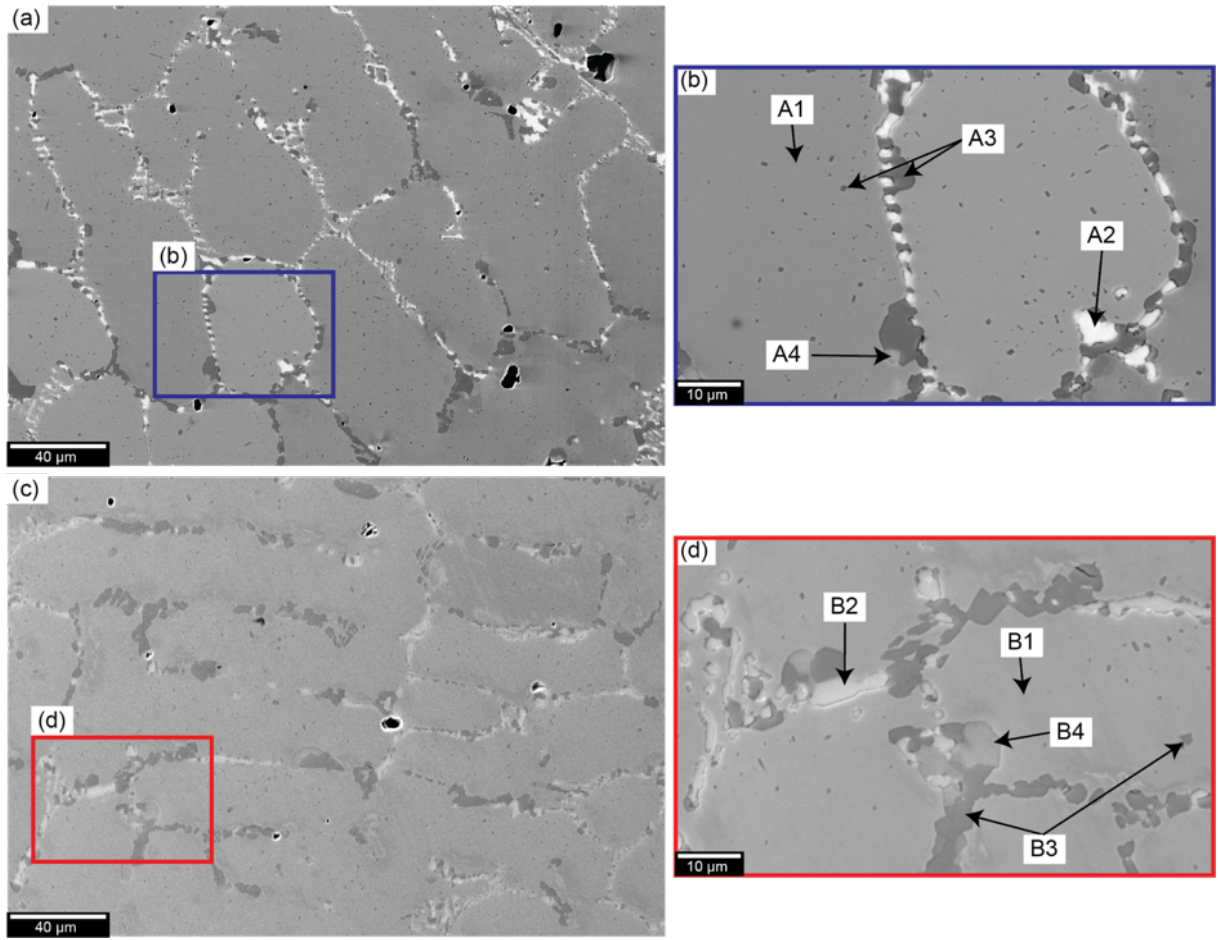

**Figure 2:** Backscatter electron images from, tube A columnar region A-ii with (a) low magnification and (b) a high magnification region, and tube B columnar region B-iv with (c) low magnification and (d) a high magnification region.

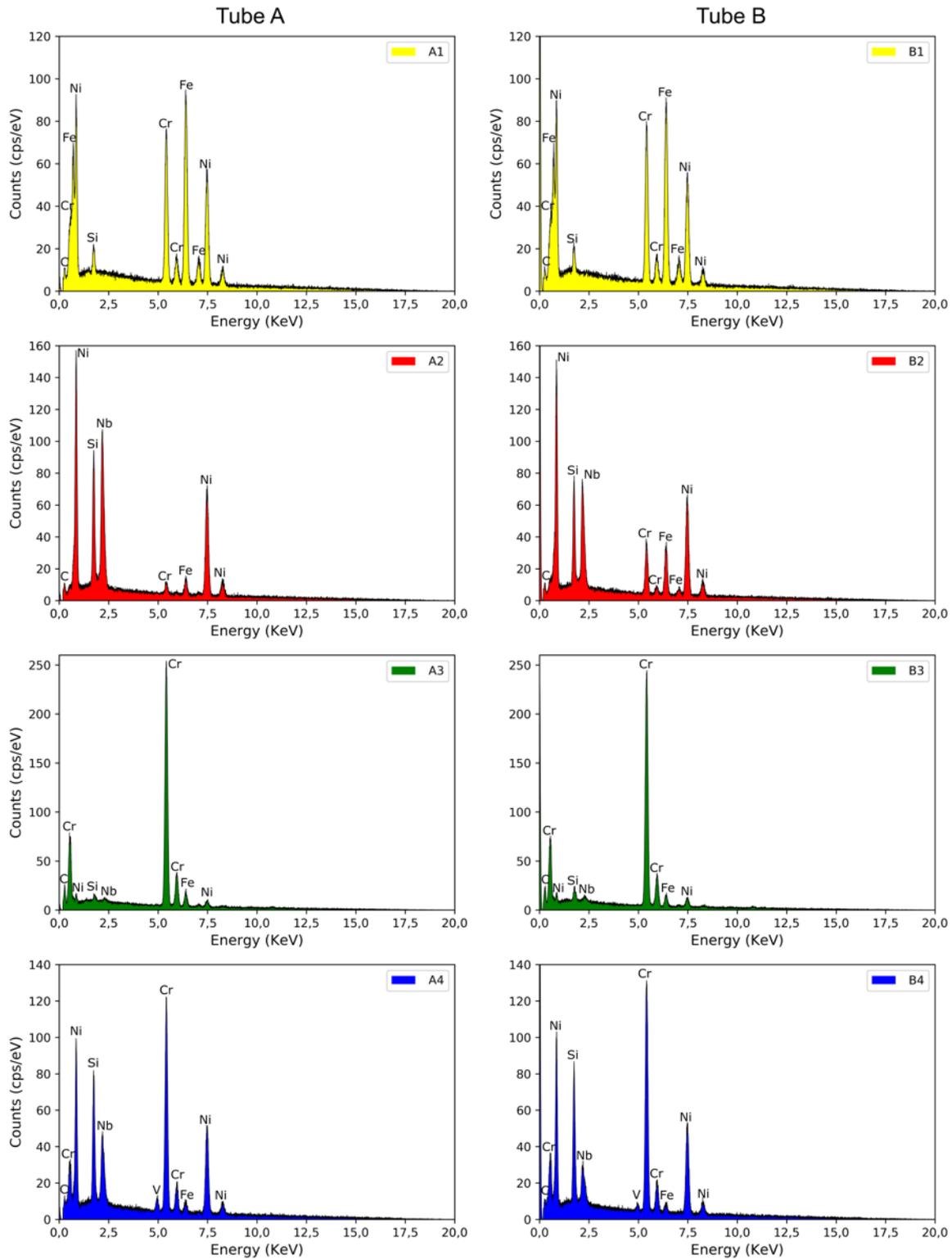

Figure 3: EDS analysis were performed on tube A on the phases noted A1, A2, A3 and A4 inside Figure 2 (b), and on tube B on the phases noted B1, B2, B3 and B4 inside Figure 2 (d).

### 3.1.3. Carbides distribution along tube section

As the backscatter electron imaging provides sufficient contrast to segment and count the area fractions of the carbide types, this was repeated on representative areas from the OD to the ID of each tube. The area fraction analysis is presented in Figure 4.

For both tubes:

- Phase 2 is evenly distributed along each tube section except the region close to the OD where volume fraction tends towards zero;
- Phase 3 seems to be less present inside the equiaxed grains region than in the columnar grains area;
- Phase 4 is mainly present at the OD. Also, it seems that the phase is more present along the tube section for the tube B than tube A where the phase volume fraction is clearly not significant.

When the phase 4 volume fraction increases, the volume fraction of the phase 2 decreases. One hypothesis is that during tubes service condition the temperature was not consistent along tube section and this is supported by Buchanan and Brear [30 and 5] who noted that there is a temperature gradient along tube section, where the OD was exposed to higher temperatures than ID. This mean that initial NbC phase present in as-cast condition was transformed to the G phase (phase noted 2). However, as the temperature was higher at the OD, subsequent G phase formation was not possible, so instead NbC transformed into $M_6C$ carbide (noted phase 4).

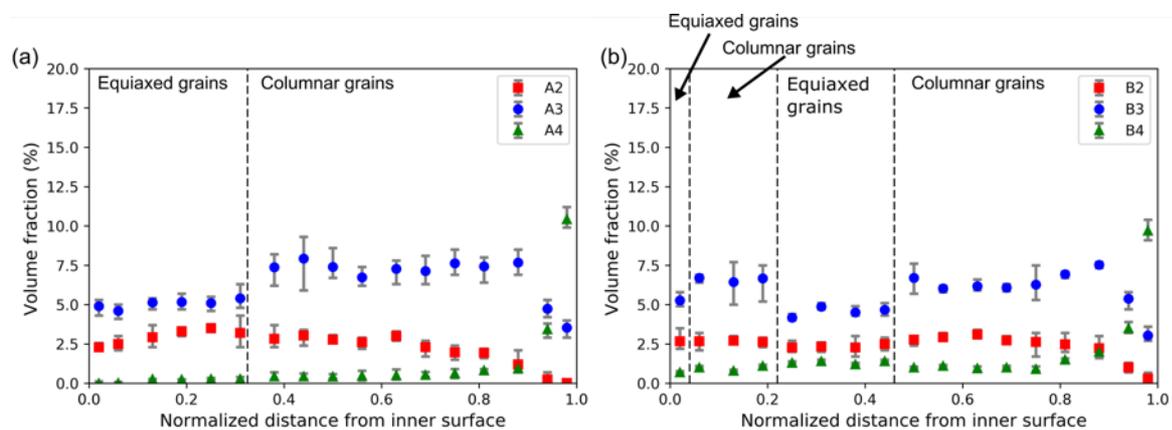

**Figure 4: Volume fraction of each carbide phases function to the distance from ID to OD for (a) tube A and (b) tube B.**

### 3.1.4. Micro cracks

During the capture of the images required to create Figure 4, microcracks were observed near the carbides in both tubes. These are shown in Figure 5, where the microcracks were observed within the regions indicated with the black dashed line in Figure 5 (a) and blue dashed line in Figure 5 (b). Micro cracks are mainly located close to the outer diameter for tube A while micro cracks are present from the middle of tube section until the OD for tube B.

Example of micro cracks are presented in Figure 5 (c) and (d) which correspond respectively to tube A and B. These two figures highlight the fact that these micro cracks are only present inside the carbide and not the matrix. The micro cracks are observed within different carbides, as Figure 5 (d) shows, with a micro crack inside a white phase in backscatter imaging, lighter than all other phases (i.e. the highest backscatter coefficient) which could corresponds to the G phase, while the other micro crack is present inside a dark phase in backscatter imaging (i.e. a lower backscatter coefficient) could possibly correspond to Cr rich carbide.

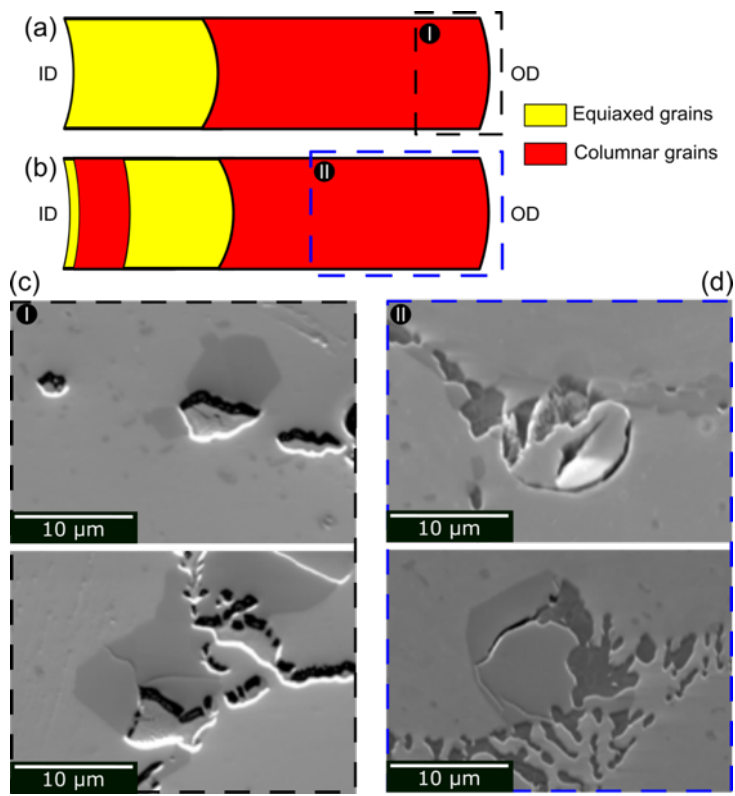

**Figure 5: Drawing of the microstructure with equiaxed and columnar grains region area of (a) tube A and (b) tube B. (c) Two SEM backscatter electron images of micro cracks inside some carbides which are present inside the I area while (d) present two SEM secondary electron images of micro cracks inside some carbides which are inside the II area.**

### 3.2. Phases identification

#### 3.2.1. Correlative EBSD-EDS phase analysis and classification

The EDS analysis reported in section 3.1.2 is insufficient to identify the phases present inside the tubes microstructure. So, to identity the phases, crystallographic information (from EBSD) was combined with chemical information (from EDS) to amplify signal-to-noise within the captured maps, the statistical approach described by McAuliffe *et al.* [32] using PCA and multivariant statistics (based upon the method developed by Wilkinson *et al.* [36]). Simulated diffraction patterns (based upon dynamical diffraction theory [37], using Bruker DynamicS) for the candidate phases were used to compare with each representative diffraction pattern using the Refined Template Matching method developed by Foden *et al.* [33].

The geometry of the pattern is calibrated using a (Ni-austenite) matrix pattern. The best phase match can be measured for each pattern and is quantified by the normalised cross correlation value, χ, calculated from a pattern projected to the refined orientation that matches the experiment.

Figure 2 to Figure 4 seem to suggest that both tubes look similar and have the same microstructure. By consequence, PCA was only performed on tube A based on the hypothesis that results will be similar for tube B.

For this analysis, an area with the matrix and all types of carbides which were previously presented was mapped and the area is shown in Figure 6. The white arrow in Figure 6 (a) indicate the expected result for each phase based on previous EDS while Figure 6 (b) shows off the crystallographic orientation of the grains by the inverse pole figure map with respect to the radial axis (i.e. the crystal directions with respect to the ID to OD axis). Figure 6 (c) present the phase map from PCA result with the classification of each phase. Note that the matrix and all the carbides are cubic (see Table 4), and so a single IPF based colour key is presented.

| Phase | Crystal System | Space Group | Lattice parameters a, b, c (nm) | | |
|---|---|---|---|---|---|
| $\gamma$ | Cubic | $Fm\bar{3}m$ | 0.35 | 0.35 | 0.35 |
| $M_{23}C_6$ | Cubic | $Fm\bar{3}m$ | 1.1 | 1.1 | 1.1 |
| G phase | Cubic | $Fm\bar{3}m$ | 1.12 | 1.12 | 1.12 |
| $M_6C$ | Cubic | $Fd\bar{3}m$ | 1.11 | 1.11 | 1.11 |

**Table 4: Synthesis of phase structure parameters inside the microstructure.**

Based on Figure 6 (c), matrix and carbides were indexed as the following results:

- Point A which corresponds to the matrix was classified as austenite. Figure 7 compare the current latent Rotated Characteristic EBSP (RC-EBSP) at this point to the other dynamics simulated pattern of different phases. The numbers above each pattern correspond to the normalised cross correlation value χ between the experimental pattern and the simulated pattern, and may be regarded as a template match quality index [33]. An index equal to 1 means that the match is extremely good, while < 0.4 corresponds to poor matching. Based on Figure 7, austenite is clearly the best match for this phase. Figure 7 (b) presents the EDS spectra at the point A while Table 5 shows the chemical composition of it. The austenite matrix is mainly composed of Cr, Fe , Ni and Si elements see also Table 5.

- Point B shows with Figure 7 and Table 5 that Cr rich was identified as $M_{23}C_6$ carbide (with M=Cr). Based on EDS results, $M_{23}C_6$ carbide presents a significant presence of Cr and limited Nb, Ni, Fe and Si elements. Figure 6 (b) also shows that $M_{23}C_6$ carbide is divided into regions of different crystal orientation.

- Points C and D based on Figure 7 correspond to the G phase carbide. Figure 6 (a) present inside the carbide a variation of contrast which may lead to a variation of the chemical composition. Therefore a point in each region was chosen for comparison. Both spectra of points C and D of Figure 7 (b) look similar with the presence of Si, Nb, Ni, Cr and limited Fe. There is a high Ni signal, but this may be a matrix contribution. However, the ratio Cr/Ni is not consistent between the two points. For the point C, the ratio is ~ 0.33 while the ratio is around ~ 0.5 for the point D. This means that it is possible to have smaller chemical variation inside the G phase carbide. Note however that carbon is difficult to measure with EDX. Similar to the $M_{23}C_6$ carbide, the G phase carbide is broken up into several regions with different crystal orientations as shown in Figure 6 (b).

- Point E was classified as $M_6C$ carbide as Figure 7 shown it. The identification of this $M_6C$ carbide is based on Figure 7 and Table 5 by the presence of significant Cr, and limited Nb, Si, Ni, Fe and V. Unlike the G phase and the $M_{23}C_6$ carbides, the $M_6C$ carbide is only composed of one single grain as shown in Figure 6 (b).

For none of the points in the maps, did the latent experimental patterns match the dynamical simulation of the NbC phase. This can be confirmed by the presence of a bright zone axis within the RC-EBSP which is not present within the NbC simulated pattern. This means that NbC carbide seems to not be present in our microstructure and that all the NbC was decomposed either into the G phase or into the $M_6C$ carbides.

| Elements | Cr | Ni | Fe | Nb | Si | Mo | Mn |
|---|---|---|---|---|---|---|---|
| A | 25 | 30.6 | 42.3 | 0 | 1.4 | 0 | 0 |
| B | 77.2 | 8.4 | 10.1 | 0.6 | 1.5 | 0.3 | 1.6 |
| C | 14 | 42 | 12.4 | 11.6 | 12.1 | 0.9 | 0.3 |
| D | 19 | 39.5 | 15.4 | 11.6 | 13 | 1.3 | 0.2 |
| E | 52.7 | 22.8 | 9.6 | 3.5 | 8.4 | 0.9 | 0.9 |

**Table 5: Composition of each measurement point (% at.) from Figure 6.**

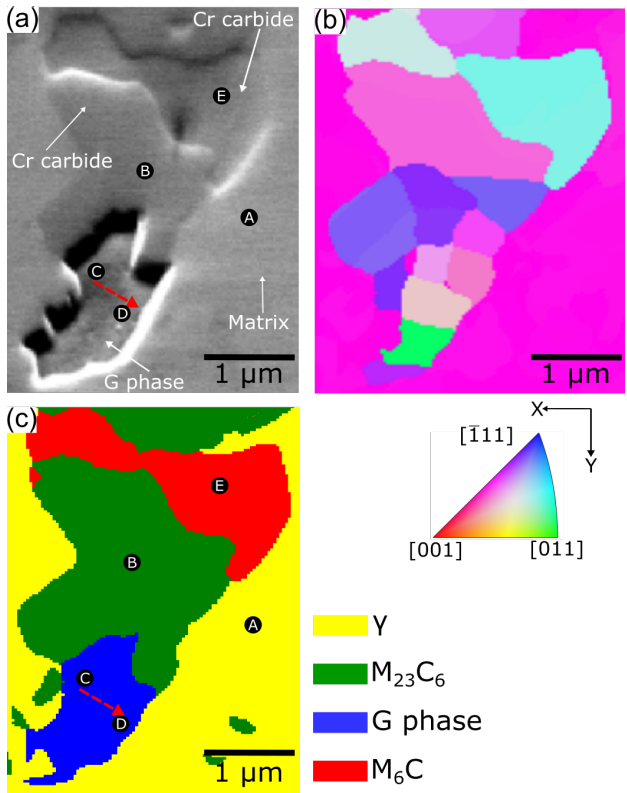

**Figure 6:** (a) SEM image of the microstructure of the tube A inside columnar grains region, (b) Inverse Pole Figure map along the X direction and (c) results from the PCA with the phase map.

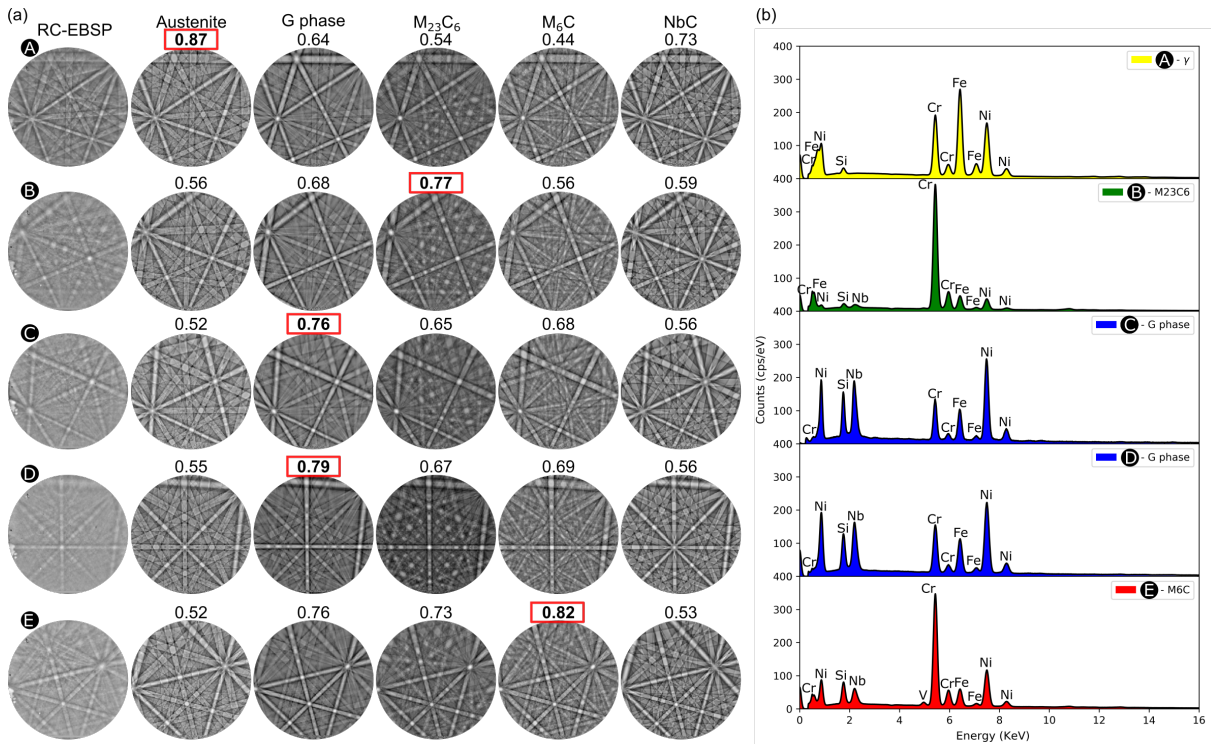

**Figure 7:** (a) Comparison between the experimental Rotated Characteristic EBSP (RC-EBSP) and the simulated phases for the points A, B, C, D and E of Figure 6. (b) EDS spectra of the points A, B, C, D and E. The numbers above each pattern are the correlation coefficients matching the simulated pattern to the RE-EBSP, and a higher number indicates a better match.

### 3.2.2. Orientation Relationship (OR)

Now that the phases have been classified and the orientations determined, it is possible to evaluate the orientation relationships between the different carbides and matrix using the data shown in Figure 6.

Table 6 presents all the OR (which are cubic-cubic pairs) found between one grain of each carbides ($M_{23}C_6$, G phase and $M_6C$) and the austenite matrix.

A maximum limit deviation of 5° between carbide direction and matrix direction was applied in order to determine the OR. The Table 6 with the superimposition of pole figure map presents with detail the results for each OR between each carbides and the matrix. Table 6 (a), (c) and (e) show the superimposition between the carbides grains and matrix pole figures map of several pairs. In each of them, the black circle indicate the location where carbide and matrix have pole deviation inferior to 5°. Table 6 (b), (d) and (f) descried how each cubic cells are connected (the cell of the carbide grain and the cell of the matrix). Moreover, in each figures, directions with small deviation between carbides and matrix are represented (matrix in black and carbide in grey).

|          | $[100]_C$/$[111]_\gamma$ | $[111]_C$/$[100]_\gamma$ | $[112]_C$/$[112]_\gamma$ | $[110]_C$/$[110]_\gamma$ | $[110]_C$/$[111]_\gamma$ | $[111]_C$/$[110]_\gamma$ | $[110]_C$/$[100]_\gamma$ | $[111]_C$/$[100]_\gamma$ |
|----------|---|---|---|---|---|---|---|---|
| $M_{23}C_6$ | OR n°1 / 3.8° | OR n°1 / 1.8° | OR n°1 / 2.9° | / | / | / | / | / |
| G phase  | / | / | / | OR n°2 / 3.8° | OR n°2 / 3.6° | OR n°2 / 2.8° | / | / |
| $M_6C$   | / | / | / | / | / | / | OR n°3 / 3.3° | OR n°3 / 3.9° |

**Table 6: OR between each carbides and the austenitic matrix. The "c" letter in the first row correspond to "carbide" and it can be M$_{23}$C$_6$, G phase or M$_6$C.**

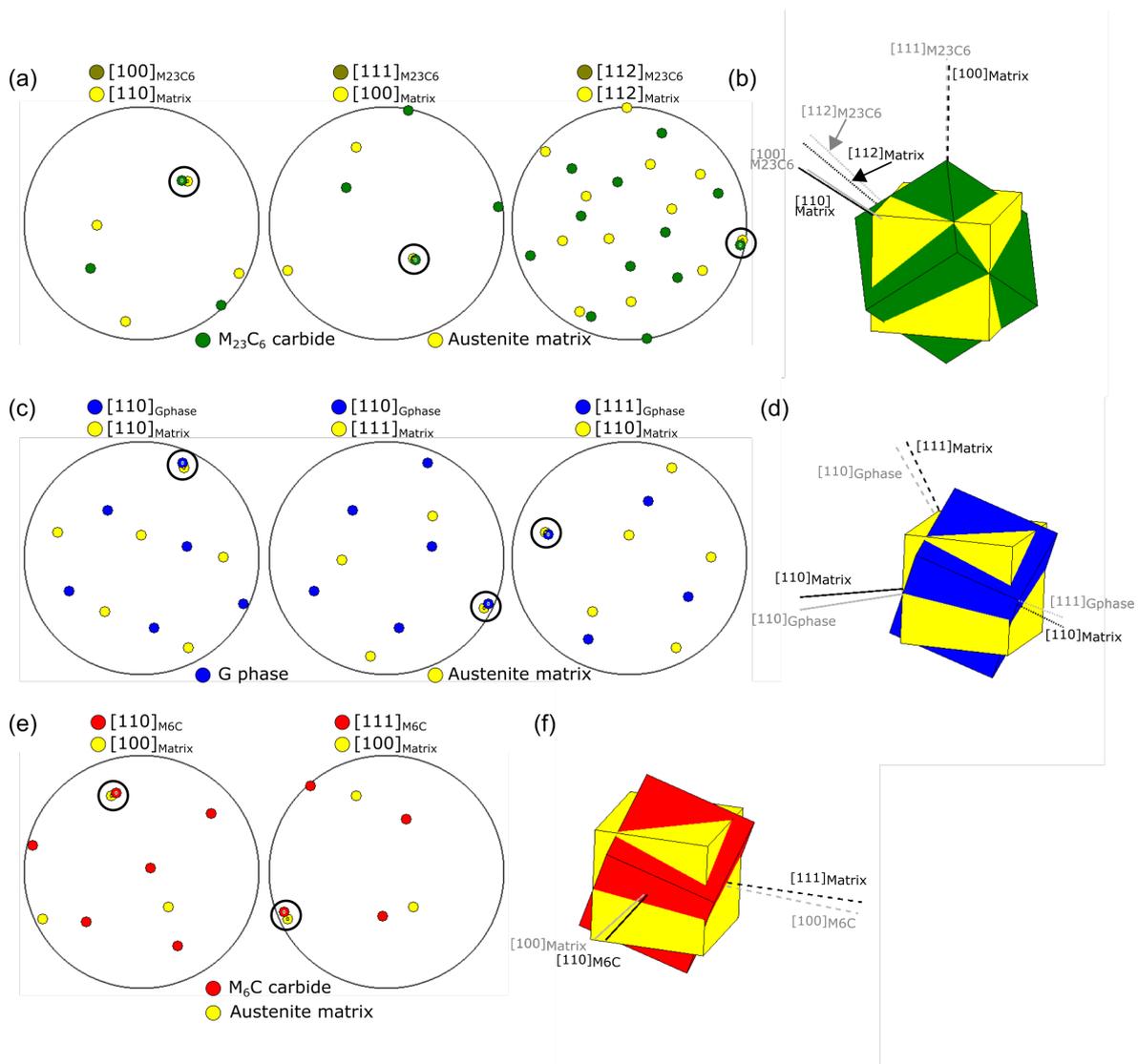

Figure 8: Superimposition of the pole figure map between (a) one $M_{23}C_6$ grain and the austenite matrix, (c) one G phase grain and the austenite matrix and (e) one $M_6C$ and the austenite matrix. Black circles show in each of them the location where carbide and matrix have pole deviation inferior to 5°. Cubic cells connection between carbides and matrix with the different direction are show for (b) the $M_{23}C_6$ grain and the austenite matrix, (d) the G phase and the austenite matrix and (f) the M6C and the austenite matrix.

### 3.2.3. Phase classification of precipitates along the tube section

To classify the precipitates along tube section and to see if there was any variation in the phases classified and their chemistry, PCA was used. Figure 9 shows all PCA results from the ID to the OD and indicates that:

- NbC phase is not present at all in our tubes (at least for these length scales) and confirms the conclusion reached in section 3.2.1. It is most likely that any NbC carbide which may have been initially present in as-cast condition were fully transformed into either the G phase or the $M_6C$ carbide during service.

- The G phase is present everywhere along the tube section, except the region close to the OD. This is consistent with Figure 4.

- In contrast, the $M_6C$ carbide is found distributed within the microstructure in the representative regions extending from the ID to the OD except in the region extremely close to the ID. The area fraction of $M_6C$ is higher closer to the OD, but these $M_6C$ carbide are also present in very small quantities for the rest of the tube section. This is consistent with Figure 4.

Figure 9 also indicates that the current results are consistent with the observation noted with the Figure 6 (c).

Buchanan *et al.* [30] observed that G phase transformation was expected to proceed preferentially at temperature below 875°C, while $M_6C$ carbide (also called η phase) is expected to preferentially occur at temperature above 920°C. In between, $M_6C$ and G phase can occur in parallel. The volume fraction of each of precipitate type will be function to the temperature exposition and chemical composition. In our case, the operating temperature is around ~ 920°C while temperature at the ID is closer to ~ 850°C. This may explain why in our case $M_6C$ occurs preferentially close to the OD while G phase is mostly present in the rest of the tube section. Buchanan *et al.* [30] also noticed that the NbC to $M_6C$ transformation seemed to be greater for NbC lamella located on the grain boundaries than NbC located on the dendrite boundaries.

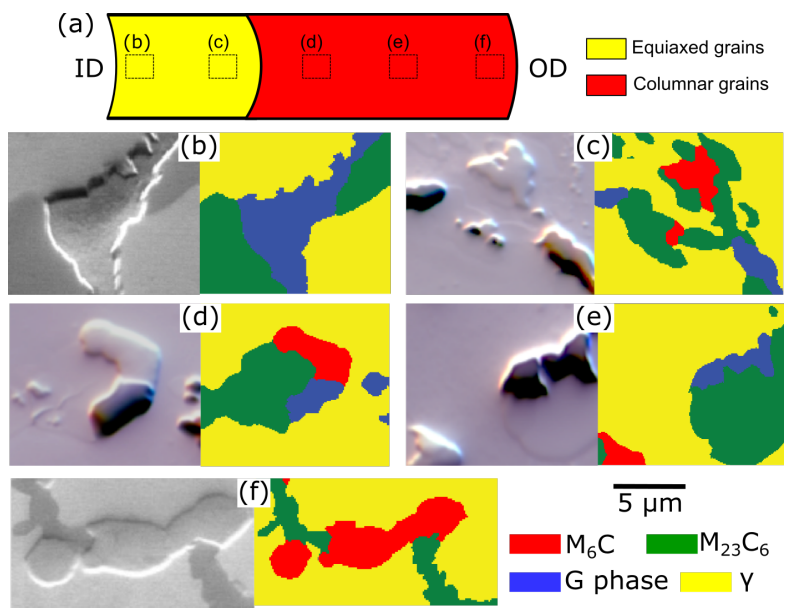

Figure 9: (a) Drawing of grain regions of tube A. (b), (c), (d), (e) and (f) correspond to several areas where PCA was performed. For each area, a secondary electron SEM image present the region of the analysis while there is also the result map with phase index.

## 4. Conclusion

In this paper, two ex-service HP40 alloy reformer tubes which have been in service under differing conditions have been compared. The aim was to assess and characterise the microstructure of both tubes with different creep life in order to highlight the effect of the accumulated creep damage by comparing their microstructure. The microstructural characterisation was performed at a range of length scales from mm to µm from the ID to the OD. A PCA approach was also used in order to classify the structure and analyse the chemistry of each phase of the microstructure.

EBSD analysis combined with optical images showed that both tubes have a microstructure with equiaxed grains close to ID and then some columnar grains towards the OD. However, tube B presents a fine band of columnar grains inside the equiaxed region. This may mean that the casting process of tube B was a little bit different compared to the tube A casting.

EDS and image segmentation suggests that both tubes are composed of the same carbides and that these carbides have a similar distribution along the tube section. Also, image analyses noticed that microcracks were present inside both tubes and they were located inside carbides. However, micro cracks were only located close to the OD for tube A while micro cracks were present from the middle to the outside of the tube wall for tube B.

PCA showed that microstructure was composed by:

- An austenitic matrix,
- A $M_{23}C_6$ carbide (with M=Cr),
- A G phase with the following composition $Ni_{16}Nb_6Si_7$,
- And a $M_6C$ carbide.

All the carbides have a cubic structure and there are orientation relationships between the carbides and the matrix phases (indicating nucleation from each other during casting). Moreover, the PCA suggested that NbC phase was not present inside our microstructure. This means that all the initial NbC phase was transformed into the G phase and $M_6C$ carbide during tube operation. Due to the elevated temperature exposure, NbC transformed into $M_6C$ at the OD, while NbC transformed into G phase in the middle and inside of the tube wall.

Phase transformation is similar for tube A and B, however, there are more micro cracks inside the tube B microstructure. This may mean that even if both tubes have accumulated creep damage, the creep damage could be a little bit more important for the tube B compared to the tube A.

## Acknowledgments

This work was financially supported by the Shell-Imperial Advanced Interfacial Materials Science (AIMS) University Technology Centre. We thank Shell Global Solutions for providing the reformer tubes and their compositions. Microscopy was conducted within the Harvey Flower Electron Microscopy Suite. TBB acknowledges funding from the Royal Academy of Engineering for his Research Fellowship. TPM was funded by EPSRC under the ACM CDT EP/21345/1 and Rolls-Royce plc. The authors also acknowledge Martin Church from Shell for the technical information on both tubes.

## Data availability

The data is available via Zenodo "DOI: http://dx.doi.org/10.5281/zenodo.4001600".